\documentstyle[12pt]{article}

\makeatletter
\if@twoside
\else \oddsidemargin 00pt \evensidemargin 00pt
\fi
\def\section{\@startsection {section}{1}{\z@}{+6.0ex plus +1ex minus
 +.2ex}{2.8ex plus .2ex}{\large\bf}}
\def\subsection{\@startsection {subsection}{2}{\z@}{+3.0ex plus +1ex
minus +.2ex}{2.3ex plus .2ex}{\normalsize\bf}}
\def\subsubsection{\@startsection{subsubsection}{3}{\z@}{+2.5ex plus
+1ex minus +.2ex}{1.5ex plus .2ex}{\normalsize\bf}}

\def\appendix{\par
 \setcounter{section}{0} \setcounter{subsection}{0}
 \def\thesection{\Alph{section}}}

\def\fo{\hbox{{1}\kern-.25em\hbox{l}}}
\def\fnote#1#2{\begingroup\def\thefootnote{#1}\footnote{#2}\addtocounter
{footnote}{-1}\endgroup}

\let\@eqnsel = \hfil

\def\eqnarray{\stepcounter{equation}\let\@currentlabel=\theequation
\global\@eqnswtrue
\global\@eqcnt\z@\tabskip\@centering\let\\=\@eqncr
$$\halign to \displaywidth\bgroup\hskip\@centering
  $\displaystyle\tabskip\z@{##}$\@eqnsel&\global\@eqcnt\@ne
  \hskip 2\arraycolsep \hfil${##}$\hfil
  &\global\@eqcnt\tw@ \hskip 2\arraycolsep $\displaystyle\tabskip\z@{##}$\hfil
   \tabskip\@centering&\llap{##}\tabskip\z@\cr}

\expandafter\ifx\csname mathrm\endcsname\relax\def\mathrm#1{{\rm #1}}\fi
\@ifundefined{mathrm}{\def\mathrm#1{{\rm #1}}}{\relax}

\makeatother

\def\beq{\begin{equation}}
\def\eeq{\end{equation}}
\def\beqra{\begin{eqnarray}}
\def\eeqra{\end{eqnarray}}
\def\bma{\begin{displaymath}}
\def\ema{\end{displaymath}}
\def\barr#1{\begin{array}{#1}}
\def\earr{\end{array}}
\def\bit{\begin{itemize}}
\def\eit{\end{itemize}}
\def\bfi{\begin{figure}}
\def\efi{\end{figure}}
\def\btab{\begin{table}}
\def\etab{\end{table}}
\def\bce{\begin{center}}
\def\ece{\end{center}}

\def\text{\textstyle}

\def\mathswitch#1{\relax\ifmmode#1\else$#1$\fi}
\newcommand{\mathswitchr}[1]
{\relax\ifmmode{\mathrm{#1}}\else$\mathrm{#1}$\fi}

\catcode`\@=11
\def\slash{\mathpalette\make@slash}
\def\make@slash#1#2{\setbox\z@\hbox{$#1#2$}%
  \hbox to 0pt{\hss$#1/$\hss\kern-\wd0}\box0}
\def\siml{\mathrel{\mathpalette\@versim<}} % approximately less
\def\simg{\mathrel{\mathpalette\@versim>}} % approximately greater
\def\@versim#1#2{\lower2.5\p@\vbox{\baselineskip\z@skip\lineskip-.2\p@
    \ialign{$\m@th#1\hfil##\hfil$\crcr#2\crcr\sim\crcr}}}

\def\haf{\frac{1}{2}}

\def\d{\partial}

\begin{document}
\thispagestyle{empty}
\null
\hfill CERN-TH.6940/93
\vskip 1cm
\vfil
\begin{center}
{\Large \bf CONDITIONS ON SUPERSYMMETRY SOFT-BREAKING TERMS FROM
GUTs
\par} \vskip 2.5em
{\large
{\sc G.F.\ Giudice\fnote{*}{\rm On leave of
absence from INFN, Sezione di Padova.}}
and {\sc E.\ Roulet} \\[1ex]
{\it Theory Division, CERN, Geneva, Switzerland} \\[2ex]
\par} \vskip 1em
\end{center} \par
\vskip 4cm
\vfil
{\bf Abstract} \par
We study the effect of integrating out the heavy fields in a
supersymmetric GUT which does not contain small mass parameters
in the limit of exact supersymmetry. The trilinear ($A$) and
bilinear ($B$) coefficients of the supersymmetry soft-breaking
terms of the low-energy effective theory are related in a simple
and model-independent way to those of the underlying theory.
{}From these relations, we obtain the bound $|B|\ge 2$, which, together
with the requirements of stability of the potential and electroweak
symmetry breaking, imposes severe constraints on the space of allowed
supersymmetric parameters. In models based on supergravity with
a flat K\"ahler metric, we obtain $B=2$, instead of the relation
$B=A-1$ usually used in phenomenological applications. The low-energy
theory contains also a supersymmetric mass term $\mu$ for the two
Higgs doublets, which is of the order of the supersymmetry-breaking
scale.
\par
\vskip 2cm
\noindent CERN-TH.6940/93 \par
\vskip .15mm
\noindent July 1993 \par
\null
\setcounter{page}{0}
\clearpage

Low-energy supersymmetric theories of fundamental interactions
contain, in addition to the parameters of the Standard Model (SM), a
set of new unknown parameters, which can only be the subject of
theoretical speculation until we are able to derive them from
experimental data. These parameters appear in the supersymmetry
soft-breaking terms and are directly related to the mechanism of
supergravity breakdown. Their typical mass scale is expected to be
${\cal{O}}(M_W)$, if supersymmetry is to solve the naturalness problem
of the SM.

Even more obscure is the origin of the $\mu$-term, the mixing-mass
of the two Higgs doublet superfields of the Minimal Supersymmetric
Standard Model (MSSM). This mass term is invariant under supersymmetry
and therefore it seems unrelated to the weak scale, although it is
phenomenologically required to be ${\cal{O}}(M_W$). It has been previously
proposed that a term $\mu\sim {\cal{O}}(M_W)$ can
be accommodated in models
with an extra gauge-singlet superfield \cite{nil} or generated as  a
supergravity-breaking effect in theories with non-trivial K\"ahler
metrics \cite{giu}, or induced by higher-dimension operators \cite{kim}.

Here we want to concentrate on models where the solution of the
$\mu$-problem lies at some Grand Unified Theory (GUT) scale. This is
of particular interest after measurements at LEP have revealed the
remarkable property of the MSSM that gauge couplings unify at a scale
$M_X\sim 10^{16}$ GeV. We will therefore consider theories with a GUT
threshold at $M_X$ and assume that, at that scale, $\mu=0$
in the absence of supersymmetry-breaking effects.
This is dictated by a naturalness criterion, since, if non-vanishing,
$\mu$ should be of the order of the only available scales, $M_X$ or
$M_{Pl}$. The relation $\mu=0$ can be the result of some symmetry at
the GUT level (as, for example, in the pseudo-Goldstone bosons model
of ref. \cite{gia}), of the specific field
content of the heavy theory (as in the
``missing partner" models \cite{dim}) or of a mere fine-tuning of the
parameters (as in the ordinary supersymmetric SU(5) model \cite{gut}).
Whatever
the reason is, very little information on the structure of the GUT
theory is needed in order to derive the form of the soft-breaking
terms of the effective low-energy theory, as was shown in
ref. \cite{joe}. By extending the analysis of ref.
\cite{joe}, we connect with simple
relations the supersymmetry-breaking parameters of the GUT with the
low-energy ones and with the induced $\mu$-term. From these
relations, some constraints on the low-energy supersymmetry-breaking
parameters can also be derived.

In the limit $M_{Pl}\to\infty$ after supergravity breaking, the theory
at the GUT scale is defined by a softly broken supersymmetric
Lagrangian. In particular, the potential for the complex scalar fields
$z$ is \cite{rot}:
\beq
V(z^*,z)=\left|{\d f\over \d z}\right|^2+m^2\left|{z}\right|^2+
m(A_Xf^{(3)}+B_Xf^{(2)}+{\mathrm {h.c.}})+{1\over 2}\sum_kD_k^2 ,
\eeq
where $D_k=z^\dagger T_{(k)}z$, $T_{(k)}$ are the gauge group generators,
$A_X$, $B_X,\ m$ are the soft-breaking parameters,
$f^{(2)}$ and $f^{(3)}$ are the terms in the superpotential respectively
bilinear and trilinear in the fields $z$. For simplicity, we assume
that the superpotential does not contain terms linear in the fields,
and thus $f=f^{(2)}+f^{(3)}$. If the underlying supergravity theory
has a flat K\"ahler metric, then \cite{rot}:
\beq
B_X=A_X-1.
\eeq

For general supergravity couplings, eq. (2) does not hold, and $m,\
A_X$ and $B_X$ are no longer universal but become matrices. This is
usually the case in theories derived from
superstrings \cite{iba}. Motivated
by the observed suppression of flavor-changing neutral currents, we
will assume universal couplings in eq. (1), allowing only for $B_X\neq
A_X-1$. We will briefly comment later on the general case.

At $M_X$ the GUT is spontaneously broken to the Standard Model.
We define $\bar{z} \sim {\cal{O}}(M_X)$ to
be the vacuum expectation values
of the fields $z$ in the limit of exact supersymmetry.
Since supersymmetry is not spontaneously broken at the
scale $M_X$:
\beq
{{\d f}\over {\d z}}(\bar{z})=D_k (\bar{z})=0.
\eeq
Since we are
assuming that all mass parameters entering the superpotential $f$ are
${\cal{O}}(M_X)$, we can distinguish the fields $z^a$ appearing in the
potential in three classes. In the limit of exact supersymmetry, $z^A$
 are complex scalars with mass ${\cal{O}}(M_X)$, $z^\alpha$ are massless
complex scalars with $\langle z\rangle=0$ (which corresponds to the
low-energy fields), and $z^K$ are real scalars belonging to the
massive vector supermultiplets of the broken generators.
We now want to
integrate out the heavy fields $z^A$ and $z^K$ in order to obtain the
low-energy effective theory, following the same procedure as in ref.
\cite{joe}\footnote{We have tacitly assumed
$M_X\ll M_{Pl}$. This is just to
simplify the calculation, but our results are valid also for $M_X\sim
M_{Pl}$, since, as shown in ref. \cite{joe}, the renormalizable
interactions of the low-energy effective theory
do not contain any power of $M_X/M_{Pl}$.}.
Using a series expansion in $m/M_X$ to solve the equations of
motion for the heavy fields, we obtain:
\beqra
z^A&=&\bar z^A+\Phi^A-(f^{-1})^{AB}\left[m\left(B^*_X+{1\over
B_X-A_X}\right) \Phi^*_B \right. \nonumber \\
&&\left.+\haf f_{BCD}\Phi^C\Phi^D+\haf
f_{B\alpha\beta}z^\alpha z^\beta\right]+{\cal {O}}\left({m^3\over
M_X^2}\right) ,\\
z^K&=&-\haf({\cal{M}}^{-1})^{KL}D_L(\Phi^A,z^\alpha )
+{\cal {O}}\left({m^3\over M_X^2}\right) ,
\eeqra
\beq
\Phi^A\equiv m(B_X-A_X)^*(f^{-1})^{AB}\bar z^*_B,\ \ \ \ \
{\cal{M}}_{KL}\equiv {\bar{z}}^\dagger \{ T_{(K)},T_{(L)}\} {\bar{z}},
\eeq
where
$f_{AB}={\d^2f\over \d z^A\d z^B}|_{z=\bar z} ,\ \
f_{abc}={\d^3f\over \d z^a\d z^b\d z^c}|_{z=\bar z}$, etc.

Plugging these expressions back in eq. (1), we obtain the potential of
the low-energy effective theory
\beq
V_{eff}=\left| {\d f_{eff}\over\d
z^\alpha}\right|^2+m^2|{z_\alpha}|^2+m(Af_{eff}^{(3)}
+Bf_{eff}^{(2)}+{\mathrm {h.c.}})+{1\over
2}\sum_{k'}D_{k'}^2,
\eeq
where the index $k'$ runs over the unbroken generators and
\beq
f_{eff}^{(3)}={1\over 6}f_{\alpha\beta\gamma}z^\alpha
z^\beta z^\gamma
,\ \ f_{eff}^{(2)}=\mu_{\alpha\beta}z^\alpha z^\beta,\ \
f_{eff}=
f_{eff}^{(3)}+f_{eff}^{(2)},
\eeq
\beq
A=A_X ,
\eeq
\beq
B=A_X-B_X+{1\over (A_X-B_X)^*} ,
\eeq
\beq
\mu_{\alpha\beta}=m(A_X-B_X)^*C_{\alpha\beta}, \ \
C_{\alpha\beta}\equiv -{1\over 2}f_{\alpha\beta A}(f^{-1})^{AB} \bar
z_B^*.
\eeq

Therefore, the potential of the low-energy theory, eq. (7), has the
same form as eq. (1), with the soft-breaking parameters related by
eqs. (9) and (10) and an induced $\mu$-term as in
eq. (11). Notice that the
case of a flat K\"ahler metric, eq. (2), corresponds to
\beq
B=2.
\eeq
For generic $A_X$ and $B_X$ we obtain from eq. (10):
\beq
|{B}|\geq 2.
\eeq
The constraints in eqs. (12) and (13) apply to
the low-energy $B$ parameter
and are relevant to phenomenological applications.

All the dependence on the details of the GUT model is contained in the
parameter $C_{\alpha\beta}$ in eq. (11). There is a class of theories
in which $C_{\alpha\beta}$ turns out to be independent of the details
of the model: the supersymmetric GUT with Higgs as pseudo-Goldstone
bosons \cite{gia}.

In these models, the Higgs sector is globally invariant under a group
$G_{gl}$ larger than the GUT gauge group $G_{loc}$. At $M_X$,
$G_{loc}$ is broken to $H_{loc}$ (the SM gauge group) and $G_{gl}$ to
$H_{gl}$. The Goldstone bosons corresponding to the broken generators
of $G_{gl}/H_{gl}$, not contained in $G_{loc}/H_{loc}$ and therefore
not eaten by the Higgs mechanism, are physical particles belonging to
massless chiral supersymmetric multiplets. After inclusion of
supersymmetry soft-breaking terms, these particles, interpreted as
the low-energy Higgs bosons, acquire masses ${\cal{O}}(m)$.
The simplest model of this kind is based on $G_{loc}=$SU(5) and
$G_{loc}=$SU(6) \cite{gia}, but
models based on $G_{loc}=$SO(10) \cite{bar} or
larger gauge groups with an automatic
$G_{gl}$ invariance \cite{ba2} have also been proposed. We want now to
compute $C_{\alpha\beta}$ in this class of models.

The invariance of the superpotential under $G_{gl}$ implies:
\beq
{\d f\over \d z^a}T^a_{(i)b}z^b=0,
\eeq
where $T^{(i)}$ are the generators of $G_{gl}$. By differentiating
eq. (14) at $z=\bar z$, we get
\beq
f_{ac}T^a_{(i)b}\bar z^b+f_aT^a_{(i)c}=0.
\eeq
Since supersymmetry is not spontaneously broken at the scale $M_X\
(f_a=0)$, any non-vanishing combination
$T^a_{(i)b}\bar z^b$ corresponds to a
massless (in the limit of exact supersymmetry) mode. We can now
construct the orthonormal Goldstone states $G^\alpha$ as follows:
\beq
G^\alpha=U^{*\alpha}_a(z^a-\bar z^a),
\eeq\beq
U^a_\alpha\equiv N_{\alpha i} T^a_{(i)b}\bar z^b,\ \ \ \
N_{\alpha i}\equiv
{V_{\alpha i} \over \sqrt{\pi_\alpha}},
\eeq
where $\pi_\alpha$ and $V_{\alpha i}$ are respectively the non-zero
eigenvalues and associated orthonormal eigenvectors of the matrix
$\Pi_{ij}\equiv \bar z^*_aT^a_{(i)b}T^b_{(j)c}\bar z^c$. These
definitions ensure that
\beq
U^{*\alpha}_aU^a_\beta=\delta^\alpha_\beta .
\eeq
By differentiating eq. (14) with respect to $z^c$ and $z^d$ at $z=\bar
z$, and then multiplying the result by $U^c_\alpha N_{\beta i}
(f^{-1})^{DE}\bar z^*_E$, we obtain from the definition of $C_{\alpha
\beta}$, eq. (11):
\beq
C_{\alpha\beta}=\haf \delta_{\alpha\beta},
\eeq
a result which is independent of the details of the GUT model.

In the case of the MSSM, $f^{(2)}_{eff}$ consists of only one term,
the two Higgs doublet mixing mass $\mu H_1H_2$,
and therefore there is only
one $C$ parameter. With the definition
\beq
H_1={1\over\sqrt{2}}(z_1+iz_2),\ \ \ H_2={1\over\sqrt{2}}(z_1-iz_2),
\eeq
we obtain, from eq. (19), $C=1$ and from eq. (11):
\beq
\mu =m(A_X-B_X)^*.
\eeq
This coincides with the relation $\mu =m$
obtained in ref. \cite{gia} in the
case of a flat K\"ahler metric.

If the low-energy Higgs doublets are not Goldstone bosons of some
global symmetry spontaneously broken at $M_X$, the parameter $C$ will
generally depend on unknown couplings of the GUT. As an example
consider the simplest GUT, SU(5), with Higgs superpotential
\beq
f={M_\Sigma\over 2}{\mathrm {Tr}}\Sigma^2+{M_H\over 2}
H_1H_2+{\lambda\over 3}
{\mathrm {Tr}}\Sigma^3+{\alpha\over 3}H_1\Sigma H_2,
\eeq
where $H_1$ ($\bar{\bf 5}$) and $H_2$ ({\bf 5}) contain the
low-energy Higgs doublets and $\Sigma$ ({\bf 24})
spontaneously breaks SU(5) into the SM:
\beqra
\langle\Sigma\rangle={M_\Sigma \over \lambda}\pmatrix{2&&&&\cr
&2&&&\cr
&&2&&\cr &&&-3&\cr &&&&-3}.
\eeqra
The condition that $\mu=0$ at $M_X$, in the absence of supersymmetry
breaking, is achieved by a fine-tuning of the parameters:
\beq
{2\alpha\over\lambda}={M_H\over M_\Sigma}.
\eeq
The parameter $C$ can be directly computed from its definition in eq.
(11) and is equal to:
\beq
C=-{\alpha\over 2\lambda}=-{M_H\over 4 M_\Sigma}.
\eeq
In the context of this minimal SU(5) model, measurements of the
low-energy parameters can give information on the coupling constants
of the GUT.

We want to stress that the relations (9)--(11) follow only from our
assumption that the low-energy fields, in the limit of exact
supersymmetry, are exactly massless at the scale $M_X$. However if the
GUT couplings do not induce a $\mu$-term, in other words if $C=0$,
then eqs. (10) and (11) do not contain any information.
This is the case, for
instance, of the GUT models with ``missing partners" \cite{dim}, where no
single term of the superpotential contains both $H_1$ and $H_2$, and
therefore $C=0$ at tree level. It is possible that loop corrections
induce a $C\neq 0$ or that the origin of the $\mu$-term in these
models is completely unrelated to the GUT scale.

If the soft-breaking masses in the second term of eq. (1) are not
universal, a new model-dependent coefficient enters in eq. (10),
generally invalidating the constraint of eq. (13). However if
deviations from universality are small, as seems to be required by
the strong observed suppression of flavor-changing neutral current
processes, our conclusion should not be drastically modified. A
non-universality of the $A$-term is important only if $H_1H_2$ is coupled
to superheavy fields in more than one term.

In the case of the MSSM, we can derive further
constraints from eqs. (9)--(11).
Since the $D$-terms for the neutral components of the
Higgs doublets vanish in the direction $|{H_1}|=|{H_2}|$, the
stability of the potential, eq. (7), requires:
\beq
m^2+|{\mu}|^2\geq |{Bm\mu}| .
\eeq
{}From eqs. (9)--(11), this implies
\beqra
\left|{\mu\over m}\right| \leq\sqrt{|C|}\leq{1\over
|{A_X-B_X}|}\ \ \ \ \ \ &&{\mathrm {if}}\ \
|C|<1\cr
{1\over |{A_X-B_X}|}\leq\sqrt{|C|}\leq \left|{\mu\over m}\right|
\ \ \ \ \ \ &&{\mathrm {if}}\ \ |C|>1
\eeqra
and any value of $A_X-B_X$ and $\mu /m$ is allowed for $|C|=1$.

One of the most attractive aspects of the MSSM is that the
renormalization of the parameters of the theory from $M_X$ to low
energies induces the breaking of the electroweak symmetry. We can
therefore investigate the constraints imposed by SU(2)$\times$U(1)
breaking with the additional relations among the parameters
that we have found here.
For a fixed value of the top-quark Yukawa coupling constant,
we run the Renormalization Group Equations (RGEs) from $M_X$ to the
low-energy scale and study the region of the soft-breaking parameters
where: $i)$ the potential is bounded from below at any energy between
the weak scale and $M_X$; $ii$) electroweak symmetry is spontaneously
broken;
$iii$) the mass spectrum of the supersymmetric
particles satisfies the present experimental bounds.  We have also
corrected the potential with the dominant one-loop contribution
coming from the top--stop sector, and we have verified the stability of
our results under variations of the scale where we stop the running of
the RGE. Figure 1 shows the allowed regions for the parameters $A$ and
$\mu/m$ for three choices of $B$: $B=2$ (flat K\"ahler metric), $B=5$,
and $B=(1+| \mu/m|^2)/(\mu/m)$ (Higgs as pseudo-Goldstone bosons).
Figure 2 shows the allowed regions for the parameters $B$ and $\mu/m$
for $A=0$ and $A=5$. The constraints of eq. (13) and eq. (26)
are also shown in fig. 2; notice that they respectively
correspond to the cases of a flat K\"ahler metric and of Higgs as
pseudo-Goldstone bosons. In figs. 1a and 2a the top-quark Yukawa
coupling constant is chosen such that $m_t=\sin\beta\cdot 140$ GeV and in
figs. 1b and 2b $m_t=\sin\beta \cdot 180$ GeV, where
$\tan\beta=v_2/v_1$, the
ratio of the two vacuum expectation values; in the MSSM, sin$\beta$ is
forced to satisfy $1/\sqrt{2}<\sin\beta<1$. These figures show the strong
existing constraints, especially for $m_t=\sin\beta\cdot 180$ GeV, where
the top-quark Yukawa is close to its infrared fixed-point.

In conclusion, we have shown that in theories where $\mu=0$ at $M_X$,
in the limit of exact supersymmetry, the GUT couplings together with
the supersymmetry-breaking effects can, and generally will, induce a
non-vanishing $\mu$-term, which turns out to be ${\cal{O}}(M_W)$. The
supersymmetry-breaking parameters of the low-energy effective theory
are simply related to those of the GUT by eqs. (9)--(11). We obtain the
model-independent bound $|B|\geq 2$, and $B=2$ in the case of flat
K\"ahler metrics. All the dependence of the GUT model in the low-energy
theory is contained in a single parameter $C$, which can be
experimentally measured, if supersymmetry is discovered. In the class
of models where the Higgs are pseudo-Goldstone bosons, $C=1$, but in
general $C$ depends on coupling constants of the GUT. The stability of
the potential and the electroweak symmetry breaking can be used to
further constrain the soft-breaking parameters.

\bigskip

We thank J.\ Louis for useful discussions.

\bigskip

\bigskip

\noindent {\large \bf Figure captions}

\bigskip

\noindent Fig. 1. The regions of parameter space $A$--$\mu / m$ allowed
by the consistency conditions {\it i)--iii)} described in the text,
for $B=2$ (flat K\"ahler metric), $B=5$, and $|B|=|\mu /m|+|m/\mu |$
(PGB -- hypothesis of Higgs as pseudo-Goldstone bosons). The
top-quark Yukawa coupling constant is chosen such that
$m_t=\sin\beta\cdot 140$ GeV (1a) and $m_t=\sin\beta\cdot 180$ GeV (1b).

\bigskip

\noindent Fig. 2. The regions of parameter space $B$--$\mu / m$ allowed
by the consistency conditions {\it i)--iii)} described in the text,
for $A=0$ and $A=5$. The constraints $|B|\ge 2$ of eq. (13) and
$|B|\le |m/\mu |+|\mu /m|$ of eq. (26) are also shown.
The top-quark Yukawa coupling constant is chosen such that
$m_t=\sin\beta\cdot 140$ GeV (2a) and $m_t=\sin\beta\cdot 180$ GeV (2b).


\begin{thebibliography}{99}
\frenchspacing
\newcommand{\np}[3]{{\sl Nucl. Phys.} {\bf B#1} (19#2) #3}
\newcommand{\pl}[3]{{\sl Phys. Lett.} {\bf B#1} (19#2) #3}
\newcommand{\pr}[3]{{\sl Phys. Rev.} {\bf D#1} (19#2) #3}
\newcommand{\pp}[3]{{\sl Phys. Rep.} {\bf #1} (19#2) #3}
\newcommand{\zp}[3]{{\sl Z. Phys.} {\bf C#1} (19#2) #3}
\newcommand{\ptp}[3]{{\sl Prog. Theor. Phys.} {\bf #1} (19#2) #3}
\newcommand{\spj}[3]{{\sl Sov. Phys. JETP} {\bf #1} (19#2) #3}


\bibitem{nil}H.P.\ Nilles, \pp{110}{84}{1}.
\bibitem{giu}G.F.\ Giudice and A.\ Masiero, \pl{206}{88}{480}.
\bibitem{kim}J.E.\ Kim and H.P.\ Nilles, \pl{138}{84}{150} and
{\bf 263} (1991) 79; \\ E.J.\ Chun, J.E.\ Kim, and H.P.\ Nilles,
\np{370}{92}{105}; \\ J.A.\ Casas and C.\ Mu\~noz, \pl{306}{93}{288}.
\bibitem{gia}K.\ Inoue, A.\ Kakuto, and H.\ Takano, \ptp{75}{86}{664};
\\ A.A.\ Anselm, \spj{67}{88}{663}.
\bibitem{dim}S.\ Dimopoulos and F.\ Wilczek, {\it Erice Summer
Lectures}, Plenum Press, New York, 1981; \\
B.\ Grinstein, \np{206}{82}{387}; \\
H.\ Georgi, \pl{108}{82}{283}; \\
A.\ Masiero, D.V.\ Nanopoulos, K.\ Tamvakis, and T.\ Yanagida,
\pl{115}{82}{380}.
\bibitem{gut}E.\ Witten, \np{188}{81}{513}; \\
N.\ Sakai, \zp{11}{81}{153}; \\
S.\ Dimopoulos and H.\ Georgi, \np{193}{81}{150}.
\bibitem{joe}L.\ Hall, J.\ Lykken, and S.\ Weinberg, \pr{10}{83}{2359}.
\bibitem{rot}R.\ Barbieri, S.\ Ferrara, and C.A.\ Savoy,
\pl{119}{82}{343}; \\ S.\ Soni and A.\ Weldon, \pl{126}{83}{215}.
\bibitem{iba}L.\ Iba\~nez and D.\ Lust, \np{382}{92}{305}; \\
V.S.\ Kaplunovsky and J.\ Louis, CERN preprint CERN-TH.6809/93 (1993).
\bibitem{bar}R.\ Barbieri, G.\ Dvali, and A.\ Strumia,
\np{391}{93}{487}.
\bibitem{ba2}R.\ Barbieri, G.\ Dvali, and M.\ Moretti, CERN preprint
CERN-TH.6840/93 (1993).

\end{thebibliography}
\end{document}